\documentclass[10pt,conference]{IEEEtran}
\IEEEoverridecommandlockouts
\usepackage{cite}
\usepackage{amsmath,amssymb,amsfonts}
\usepackage{graphicx}
\usepackage{textcomp}
\usepackage{verbatim}
\usepackage[utf8]{inputenc}
\usepackage{svg}
\usepackage{tikz}
\usepackage{subfig}
\usepackage{cite}
\usepackage{float}
\usepackage{booktabs}
\usepackage{comment}
\usepackage{multirow}
\usepackage{makecell}
\usepackage{threeparttable}
\usepackage{pgfplots }
\usepackage[english]{babel}
\usepackage{amsmath}
\usepackage{graphicx,url}
\usepackage{algorithm}
\usepackage{algpseudocode}
\usepackage{algorithmicx}

\usepackage[table,xcdraw]{xcolor}

\def\BibTeX{{\rm B\kern-.05em{\sc i\kern-.025em b}\kern-.08em
   T\kern-.1667em\lower.7ex\hbox{E}\kern-.125emX}}

\ifCLASSINFOpdf
\else
\fi

\begin{document}

\title{Deadline-Aware Multi-Agent Reinforcement Learning for TSN-Based Vehicular Edge Networks}

\author{\IEEEauthorblockN{
Bernardo A. C. Pereira\IEEEauthorrefmark{1},
Marcos Carvalho\IEEEauthorrefmark{1},
Fatih Temiz\IEEEauthorrefmark{2},
Shavbo Salehi\IEEEauthorrefmark{2},
Melike Erol-Kantarci\IEEEauthorrefmark{2}, Fellow, IEEE,\\
Andreas Gavrielides\IEEEauthorrefmark{3},
Johann M. Marquez-Barja\IEEEauthorrefmark{3},
Daniel F. Macedo\IEEEauthorrefmark{1}
% \IEEEauthorrefmark{5}
}
\IEEEauthorblockA{\IEEEauthorrefmark{1}Universidade Federal de Minas Gerais, Brazil\\\IEEEauthorrefmark{2}School of Electrical Engineering and Computer Science, University of Ottawa, Ottawa, Canada \\
\IEEEauthorrefmark{3}University of Antwerp - imec, IDLab - Faculty of Applied Engineering, Belgium\\
E-mails: bernardoalvescosta10@gmail.com,\{marcoscarvalho, damacedo\}@dcc.ufmg.br}\{ftemi033, ssale038, melike.erolkantarci\}@uottawa.ca\\
\{andreas.gavrielides, johann.marquez-barja\}@imec.be
}

% \author{
% \IEEEauthorblockN{
% {Bernardo A. C. Pereira$^1$},
% {Marcos Carvalho$^1$},
% {Fatih Temiz$^2$}

% {Andreas Gavrielides$^3$},
% {Daniel F. Macedo$^1$}\\
% \\
% }
% \IEEEauthorblockA{
% $~^1$Universidade Federal de Minas Gerais (UFMG, Brazil)
% \\
% $~^2$IMEC, Belgium\\
% {bernardoalvescosta10@gmail.com, \{marcoscarvalho, damacedo\}@dcc.ufmg.br}, andreas.gavrielides@imec.be\\
% \\
% }
% \\[-1.5cm]

% }

% \author{\IEEEauthorblockN{Gabriel H. M. Mato}
% \IEEEauthorblockA{\textit{Department of Electrical Engineering}\\
% \textit{Universidade Federal de Minas Gerais}\\
% Belo Horizonte, Brazl \\
% gabrielhm1@ufmg.br}
% \and
% \IEEEauthorblockN{Marcos Carvalho}
% \IEEEauthorblockA{\textit{Computer Science Department}\\
% \textit{Universidade Federal de Minas Gerais}\\
% Belo Horizonte, Brazl \\
% marcoscarvalho@dcc.ufmg.br}
% \and
% \IEEEauthorblockN{ Daniel F. Macedo}
% \IEEEauthorblockA{\textit{Computer Science Department}\\
% \textit{Universidade Federal de Minas Gerais}\\
% Belo Horizonte, Brazl \\
% damacedo@dcc.ufmg.br}
% }

% % The paper headers
% \markboth{2023 IEEE Latin-American Conference on Communications (LATINCOM)}%
% {Oliveira \MakeLowercase{\textit{et al.}}: Data consumption and user experience in video lecture sessions via Mobile Telephony Network}

\maketitle

\begin{abstract}
Vehicular edge computing (VEC) enables latency-sensitive applications by bringing computing and networking resources closer to vehicles. However, existing approaches often overlook network contention among co-located services with heterogeneous and dynamic latency requirements. While time-sensitive networking (TSN) provides bounded-latency communication, conventional and reinforcement learning-based schedulers struggle to adapt to highly dynamic vehicular environments and inter-queue dependencies. To address these limitations, we propose a multi-agent reinforcement learning (MARL) approach for queue-level scheduling in TSN-enabled VEC. Each TSN queue is assigned an autonomous agent that jointly learns the queue service order and time-slot duration to minimize deadline misses under speed-dependent latency requirements. We employ multi-agent proximal policy optimization (MAPPO) to enable coordinated yet autonomous scheduling decisions. Evaluation against single-agent, multi-agent, and non-learning-based baselines shows that MAPPO provides robust performance across different traffic profiles. Compared with centralized single-agent methods, it reduces service latency by up to 66.2\% and improves reliability by up to 271.8\%. Furthermore, unlike urgency-based heuristics, MAPPO ensures balanced scheduling while achieving lower inference times compared to other MARL methods.

\end{abstract}

\begin{IEEEkeywords}
Extended Reality, Multi-agent Reinforcement Learning, Time-Sensitive Networking, Vehicular Edge Computing
\end{IEEEkeywords}

\IEEEpeerreviewmaketitle

\section{Introduction}
\label{sec:introduction}

Vehicular edge computing (VEC) has emerged as a key infrastructure for enabling a wide range of vehicular applications, including autonomous driving support and immersive applications such as augmented and virtual reality (AR/VR)~\cite{feng2023resource}. These emerging applications introduce stringent quality of service requirements, demanding ultra-low latency and high reliability to ensure responsive operation. To meet these requirements, VEC deploys computing and networking resources at roadside units (RSUs), bringing them closer to vehicles~\cite{barakabitze2020}. In this architecture, RSUs can cache and deliver application content locally, improving service responsiveness. While existing VEC research primarily focuses on managing computational resources at edge servers, it generally assumes that edge servers operate under a best-effort Ethernet communication model. %~\cite{gao2026gspm}. 
%Recent studies show that optimizing latency only at intermediate network devices is insufficient, as host delays can significantly degrade application performance~\cite{xue2026keepon}.

Time-sensitive networking (TSN) has become a key technology for supporting latency-sensitive services over Ethernet networks. Developed by the IEEE 802.1 working group, TSN comprises a set of standards that provide bounded latency and high reliability for time-critical services~\cite{farkas2019}. Among these mechanisms, the IEEE 802.1Qbv Time-Aware Shaper (TAS) enables time-based traffic scheduling, allowing the management of co-located latency-sensitive flows~\cite{carvalho2025performance}. Nevertheless, enabling the management of mixed deadline-critical traffic in highly dynamic edge environments remains an open challenge~\cite{gavriluct2020traffic}. This challenge becomes even more complex in vehicular networks due to %fluctuations in wireless channel conditions, which increase connection instability and the risk of transmission failures caused by 
the high mobility of vehicles. To mitigate this issue, we argue that the latency budget for critical services at the edge server must be tightened as vehicle speeds increase, ensuring that packets can be successfully delivered before connectivity degradation occurs.

In this context, conventional TSN schedulers based on optimization techniques, such as satisfiability modulo theories and integer linear programming, face challenges in rapidly adapting schedules to highly dynamic environments~\cite{stuber2023survey}. Conversely, reinforcement learning (RL) has been used for TSN scheduling by enabling adaptive decision-making through continuous interaction with the environment. Nevertheless, existing studies primarily optimize intermediate network devices while overlooking the traffic sources. Furthermore, existing RL-based approaches rely on a centralized scheduling policy that determines the transmission schedule for all queues, rather than allowing each queue to make autonomous scheduling decisions~\cite{zhou2026convergence}. As a result, they are limited in capturing inter-queue dependencies in dynamic environments, particularly when latency deadlines vary over time.

\begin{figure*}[t]
  \centering
  \includegraphics[width=0.65\textwidth]{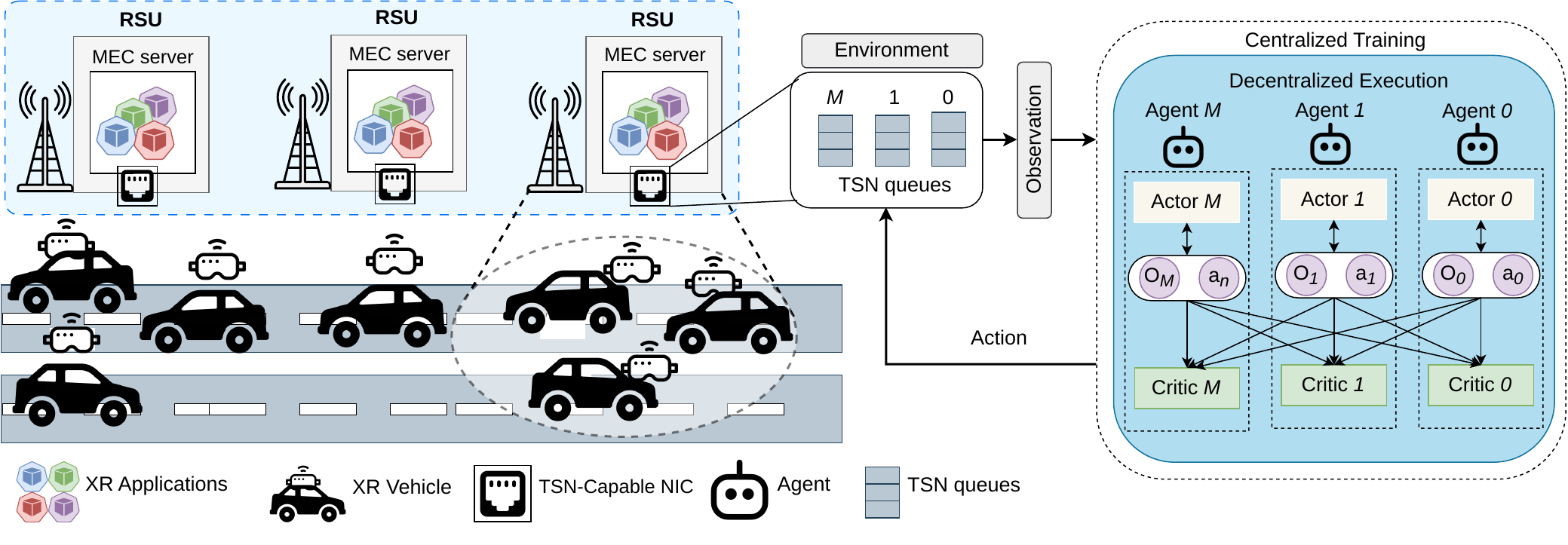}
  \caption{Overall system model.}
  \label{fig:architecture}
\end{figure*}

To address these limitations, inspired by the approach in~\cite{carvalho2026multi}, we propose a multi-agent reinforcement learning (MARL) approach in which each TSN queue is assigned an autonomous agent. The agents learn scheduling policies that jointly determine the service order and time-slot duration of each queue to minimize deadline misses. We also leverage multi-agent proximal policy optimization (MAPPO)~\cite{yu2022surprising}, which employs centralized training with decentralized execution (CTDE). During training, the centralized critic exploits global state information to produce more informative policy updates, encouraging coordinated scheduling decisions while enabling decentralized execution. We evaluate the effectiveness of the proposed MARL approach against centralized single-agent and non-learning-based baselines. The results demonstrate that MAPPO outperforms the considered approaches, reducing service latency by up to 66.2\% and improving reliability by up to 271.8\%, while achieving lower inference times than the other MARL methods.

% DRIVE: Dynamic Resource-aware Intelligent Task
% Offloading in Vehicular Edge Computing

\section{Related Works}\label{sec:related_workd}
The authors in~\cite{roberty2024configuring} employ proximal policy optimization (PPO) to dynamically optimize scheduling parameters, thereby reducing end-to-end latency in industrial network environments. Despite its effectiveness in ensuring bounded latency, the proposed approach is evaluated in a limited scenario with only two queues. This limitation may hinder its scalability and effectiveness in more complex network environments, where scheduling decisions for one queue can directly influence the performance of others. Conversely, the authors in~\cite{zhou2021mitigation} introduce a deep deterministic policy gradient (DDPG)-based method for adapting time-slot durations across multiple queues, aiming to mitigate the impact of transmission failures caused by external factors (i.e., time-synchronization errors) on application latency. Although the approach ensures bounded latency for critical applications, it does not explicitly capture the interdependence among queues, which may limit its ability to coordinate resource allocation in highly dynamic network environments with mixed and heterogeneous latency requirements. Recent studies have explored MARL solutions with different agent designs to address the complexity of TSN scheduling. For instance,~\cite{guo2024towards} divides the scheduling problem into dedicated routing and timing agents, enabling adaptive traffic management at the device level. Likewise,~\cite{garcia2025multi} introduces a multi-agent architecture within a centralized control framework, in which routing and scheduling agents collaborate to minimize latency in dynamic network environments. However, these approaches focus on network-level coordination rather than cooperative optimization of individual host-level queues. Our work learns coordinated scheduling policies that account for queue states, heterogeneous latency requirements, and dynamic edge traffic.

\section{System Model and Problem Formulation}

As illustrated in Figure \ref{fig:architecture}, we consider RSUs co-located with MEC servers and deployed along a highway segment. Let $\mathcal{U} = \{u_1, u_2, \dots, u_N\}$ denote the set of $N$ vehicles within the coverage range of the RSU, and let $\mathcal{V} = \{v_1, v_2, \dots, v_N\}$ denote their respective speeds. The MEC server hosts two classes of applications: traditional AR and semantic AR (SeAR) applications~\cite{morin2023extended}. Unlike traditional AR, which transmits the entire video content, SeAR transmits compact data units containing masks that capture the semantic content of the video. This allows SeAR to convey the relevant semantic information while reducing the amount of data transmitted over the network. Furthermore, to ensure strict temporal isolation, the server is equipped with a TSN-capable network interface card (NIC) with a set of queues, denoted as $\mathcal{Q} = \{q_{0}, q_{1}, \dots, q_{M}\}$, where each XR application is mapped to a dedicated queue. Since our focus is on queue-level scheduling at the MEC layer, we model only the downstream traffic generated by the considered applications at the edge server.

\subsection{XR Traffic, Latency and Dynamic Deadline Model}
\paragraph{XR Traffic} We adopt the statistical traffic model for XR applications presented in~\cite{morin2023extended} and incorporate the protocol data unit (PDU)-set concept as in~\cite{paymard2024pdu}. 
The traffic stream for each vehicle $u_n \in \mathcal{U}$ is modeled as a sequence of PDU-sets, where each PDU-set is indexed by $f \in \mathcal{F}_{u_n}$ and consists of a group of closely related PDUs that carry a single application-level payload. The individual PDUs within a given PDU-set are indexed by $j \in \mathcal{J}_{u_{n},f}$. The temporal characteristics of the XR traffic are determined by the arrival times of PDU-sets and their constituent PDUs. Let $t_{u_{n},f,q_{m}}$ denote the arrival timestamp of PDU-set $f$ at TSN queue $q_{m}$ of vehicle $u_{n}$. Each PDU-set consists of $N_{\text{IP}}^f$ PDUs, where $t_{u_{n},f_,j,q_{m}}$ denotes the arrival timestamp of PDU $j$ within PDU-set $f$ at the queue $q_{m}$. The PDUs are enqueued at the TSN queue according to their arrival times and served following a first-in, first-out (FIFO) policy during each transmission opportunity.  

\paragraph{Latency Model} PDUs experience queueing latency while waiting for a transmission opportunity. Let $t^s_{u_n,f,j,q_m}$ denote the service start time of PDU $j$ belonging to PDU-set $f$ of vehicle $u_n$ at queue $q_m$. To capture the worst-case waiting time at each queue, we define the age of the oldest PDU (AOPDU), representing the waiting time of the oldest PDU currently in queue $q_m$ at observation time step $t$, denoted by $g_m(t)$. Let $\mathcal{B}_m(t)$ denote the set of PDUs currently waiting in queue $q_m$. The AOPDU is given by
\begin{equation}
    g_m(t) =
    \begin{cases}
        \displaystyle \max_{j \in \mathcal{B}_m(t)}
        \left(t - t_{u_n,f,j,q_m}\right), & \text{if } \mathcal{B}_m(t) \neq \emptyset,\\[6pt]
        0, & \text{otherwise}.
    \end{cases}
\end{equation}

\noindent Due to the interdependencies among the PDUs within a PDU-set, the delayed delivery of a single PDU may prevent the timely reconstruction of an XR video frame or semantic mask. The total latency experienced by PDU-set $f$, denoted by $\Gamma_{u_n,f, q_{m}}$, is defined as the time elapsed between the arrival of its first PDU at the queue $q_{m}$ and the completion of its last PDU, expressed as $\Gamma_{u_n,f,q_{m}} = t^c_{u_n,f,q_{m}} - t_{u_n,f,1,q_{m}}$, where $t^c_{u_n,f, q_{m}}$ denotes the completion time of the last PDU in PDU-set $f$, and $t_{u_n,f,1,q_m}$ denotes the arrival time of its first PDU at queue $q_m$.

\paragraph{Dynamic PDU-Set Deadline Model} The sequence of PDU-sets $\mathcal{F}_{u_n}$ requested by vehicle $u_n$ is mapped to a tuple $\Phi_{u_n,f} = \langle src_{u_n}, dst_{u_n}, N_{\text{IP}}^f, q_m, D_{u_n,f} \rangle$, where $src_{u_n}$ and $dst_{u_n}=u_n$ denote the source RSU and destination vehicle, respectively, $N_{\text{IP}}^f$ is the number of PDUs in PDU-set $f$, $q_m$ is the associated TSN queue, and $D_{u_n,f}$ is its dynamic deadline. The deadline is defined as $D_{u_n,f} = \alpha D_{\text{base}}$, where $D_{\text{base}}$ is the fixed base deadline of the corresponding XR application. For traditional AR applications, $D_{\text{base}}$ is determined by the generation rate; for example, a 90 Hz application has a base deadline of $11.11$ ms. For SeAR applications, we consider a tighter base deadline of $2$ ms due to the importance of timely semantic information. The speed-dependent scaling factor $\alpha$ is given by
\begin{equation}
\alpha = \alpha_{\max} -
\frac{v_{avg}-v_{avg}^{\min}}{v_{avg}^{\max}-v_{avg}^{\min}}
(\alpha_{\max}-\alpha_{\min}).
\end{equation}

\noindent where $\alpha_{\max}$ and $\alpha_{\min}$ are the maximum and minimum deadline adjustment factors, respectively, $v_{avg}$ denotes the average speed of $\mathcal{U}$, and $v_{avg}^{\min}$ and $v_{avg}^{\max}$ denote the minimum and maximum average speeds considered. Thus, $\alpha$ dynamically adjusts the application deadline according to vehicle speed. As speed increases, the vehicle remains within the RSU coverage area for less time, reducing the time available for transmission. To satisfy the XR applications' requirements, the total latency of each PDU-set must not exceed its corresponding dynamic deadline $D_{u_n,f}$. Accordingly, we define the indicator of a successful on-time transmission of PDU-set $f$ from vehicle $u_n$ as
\begin{equation}
    \mathbb{I}_{u_n,f} = 
    \begin{cases} 
    1, & \text{if } \Gamma_{u_n,f,q_{m}} \leq D_{u_n,f},\\
    0, & \text{otherwise}.
    \end{cases}
    \label{eq:reliability}
\end{equation}

\subsection{TSN Schedule Model}
The network execution time is divided into fixed-duration scheduling cycles, indexed by $c$, with $\Delta t_{\text{cycle}}=1$ ms. At the beginning of each cycle, a schedule $S_c$ partitions the cycle into $M$ time slots, where $w_{c,m}$ denotes the duration (in $\mu$s) allocated to queue $q_m$. The allocated durations satisfy $\sum_{m=0}^{M} w_{c,m}=\Delta t_{\text{cycle}}$. The second step consists of determining the service order of the queues within the cycle. To govern this, we introduce a continuous priority score $p_{c,k} \in [0, 1]$. Queues are scheduled in descending order of their assigned priority score $p_{c,m}$. Furthermore, two or more queues may be assigned the same score, indicating equal urgency. In such cases, ties are resolved by granting higher precedence to queues with larger indices (i.e., $q_{M} > q_{M-1} > \dots > q_0$), reflecting standard hardware driver-level queue prioritization. Consequently, the resulting service order is defined as a permutation $\pi_c : \{1,\dots,M\} \to \{0,\dots,M-1\}$, where $\pi_c(k)$ denotes the queue index placed in the $k$-th position of the schedule. This is obtained by sorting the queues in descending order of $p_{c,m}$, and breaking ties by selecting the largest index $m$, as $\pi_c = \operatorname{argsort}_{m \in \{0, \dots, M\}} (-p_{c,m}, -m)$. Given the formalization above, the final schedule $S_c$ is defined as an ordered sequence of tuples, given by $S_c = \langle (w_{c,\pi_c(1)}, \pi_c(1)), \dots, (w_{c,\pi_c(M)}, \pi_c(M)) \rangle$, where $w_{c,\pi_c(m)}$ is the corresponding assigned time slice duration and $\pi_c(k)$ denotes the index of the queue scheduled in the $k$-th position of the cycle.

\subsection{Problem Formulation}

%In this work, we assume a strict latency constraint scenario: all PDUs belonging to an XR PDU-set must be fully transmitted prior to the dynamic deadline $D_{n,f}$. While SeAR drastically optimizes bandwidth by transmitting only essential semantic features and returning lightweight masks, the successful rendering of the user's scene is highly dependent on this temporally sensitive data. Therefore, SeAR flows needs a tighter delay budget compared to the relatively more tolerant constraints of fully offloaded traditional AR. 
The primary objective of the scheduler is to maximize the aggregate number of on-time PDU-sets across all co-located, mixed-criticality XR applications. Let $\mathbf{W}$ denote the set of time slot duration decisions and $\mathbf{P}$ denote the set of priority scores. %Here, $w_{c,k}$ and $p_{c,k}$ denote the time slice and priority score assigned to queue $q_m$ during cycle $c$, respectively. 
The scheduling problem is formulated as
\begin{align}
\max_{\mathbf{W}, \mathbf{P}} \quad & \sum_{u \in \mathcal{U}} \sum_{f \in \mathcal{F}_u} \mathbb{I}_{u,f} \label{eq:opt_obj} \\
\text{s.t.} \quad & \sum_{m=1}^{M} w_{c,m} = \Delta t_{\text{cycle}}, \quad \forall c, \label{eq:opt_cycle} \\
%& B_{tx}(w_{c,m}) \le \text{Cap}(w_{c,m}), \quad \forall c, \forall m, \label{eq:opt_cap} \\
& D_{u,f} \le D_{\text{base}}, \quad \forall u, \forall f, \label{eq:opt_ifi} \\
& D_{u,f}^{\text{SeAR}} < D_{u,f}^{\text{AR}}, \quad \forall u. \label{eq:opt_mixed}
\end{align}
 
\noindent Constraint \eqref{eq:opt_cycle} guarantees that the sum of all allocated time slots equals the predefined cycle duration $\Delta t_{\text{cycle}}$. %Constraint \eqref{eq:opt_cap} ensures that the actual bytes transmitted during time slot $w_{c,k}$, $B_{tx}(w_{c,k})$, do not exceed $\text{Cap}(w_{c,k}) = r \cdot w_{c,k}$, the maximum number of bytes transmissible during that time slot at link rate $r$. 
Constraint \eqref{eq:opt_ifi} dictates that the dynamic deadline $D_{u_{n},f}$ of any application cannot exceed its base deadline. Finally, constraint \eqref{eq:opt_mixed} strictly enforces the tighter delay requirements for SeAR applications compared to traditional AR applications.

\subsection{MDP Formulation}

We formulate the TSN scheduling problem as a multi-agent partially observable Markov decision process. Since each agent controls exactly one queue, we index agents by their queue index $m \in \{0,\dots,M\}$. Below, we describe the three essential components of each agent.
\begin{itemize}
    \item \textit{Observation Space:} At the start of scheduling cycle $c$, each queue agent $m$ receives a local observation $o_{c,m}$ that captures the traffic state and queue dynamics, given by $o_{c,m} = \langle B_{c,m}, \mathrm{M}_{c,m}, g_{c,m}, d_{c,m} \rangle$, where $B_{c,m}$ represents the backlog, defined as the total number of PDUs currently waiting for transmission in queue $q_m$. $M_{c,m}$ denotes the number of missed PDU-sets waiting in queue $q_m$. $g_{c,m}$ represents the AOPDU waiting time in queue $q_m$, while $d_{c,m}$ denotes the urgency rate, defined as $d_{c,m} = \frac{g_{c,m}}{D_{u_n,f}}$.

    \item \textit{Action Space:} At each scheduling cycle $c$, each agent $m$ generates a two-dimensional action vector. Specifically, the action for agent $m$ is defined as $a_{c,m} = \langle a_{c,m}^{dur}, a_{c,m}^{ord} \rangle$, where $a_{c,m}^{dur} \in [0, 1]$ represents the agent's desired fraction of the time slice duration. The actual time slice duration $w_{c,m}$ (in $\mu$s) allocated to queue $q_m$ is computed as $ w_{c,m} = \frac{a_{c,m}^{dur}}{\sum_{m=0}^{M-1} a_{c,m}^{dur}} \cdot \Delta t_{\text{cycle}}$. In contrast, $a_{c,m}^{ord} \in [0, 1]$ determines the service order $\pi_c(m)$ of the queue within the scheduling cycle. %The environment aggregates these scores across all agents and determines the final execution sequence $\pi_c$ by sorting the queues in descending order of their priority requests, using the hardware-level ordering for tie-breaking.

    \item \textit{Reward Function:}
    The reward is driven by the urgency rate $d_{c,m}$, which measures the AOPDU relative to its dynamic deadline, and by PDU-set deadline satisfaction. While eq.~\ref{eq:opt_obj} directly captures the objective of maximizing PDU-set deadline satisfaction through the binary indicator in eq.~\ref{eq:reliability}, this indicator provides limited intermediate feedback for RL-based scheduling. Therefore, Eq.~(8) uses a dense, bottleneck-aware surrogate reward based on queue urgency and accumulated deadline violations, providing continuous feedback to guide the policy toward higher PDU-set deadline satisfaction:
\begin{equation}
    \Psi_{c,m} = 
    \begin{cases} 
    d_{c,m}, & \text{if } d_{c,m} \leq 1 \\ 
    1 + \lambda \cdot M_{c,m} \cdot (d_{c,m} - 1), & \text{if } d_{c,m} > 1 
    \end{cases}
\end{equation}

\noindent where $\lambda$ is a scaling factor. When a queue operates safely by transmitting each PDU within its dynamic deadline ($d_{c,m} \leq 1$), the penalty simply reflects its current urgency rate. However, once a deadline is violated ($d_{c,m} > 1$), the penalty is amplified proportionally to both the violation magnitude and the number of missed PDU-sets $M_{c,m}$. %\carvalho{It should be noted that missed deadlines are calculated using eq.~\ref{eq:reliability} and incorporated into the objective function.}
%This mechanism aims to severely punish agents that consistently fail to meet application requirements, effectively mapping PDU-set failures to a high penalty. 
The global team reward $R_c$ is determined by the bottleneck queue, defined as the negative maximum penalty across all TSN queues, given by $R_c = -\max_{m \in \{0,\dots,M\}} \Psi_{c,m}$.

%By penalizing the entire team based on the worst-performing queue, this reward function implicitly forces the agents to coordinate their time slice requests and priorities, ensuring fairness and minimizing the worst-case XR application delivery failures.
\end{itemize}
 
\section{MAPPO for TSN-Based VEC}
\subsection{MAPPO Framework}
To efficiently address the optimization problem, we employ the MAPPO algorithm~\cite{yu2022surprising}. As illustrated in fig.~\ref{fig:architecture}, $M$ cooperative agents control an individual TSN queue. MAPPO extends the single-agent PPO framework to the multi-agent setting by adopting a CTDE paradigm. Specifically, each agent is equipped with an actor policy that determines its local scheduling action, while a centralized critic estimates the value of the global system state during training. The critic network has access to global information, including the observations and actions of all agents, allowing it to capture the inter-queue dependencies and coupling among the queues. In contrast, each actor receives only its corresponding local observation and learns a policy for selecting its action. The centralized critic network provides a more informative value estimate for computing the policy gradient and advantage function, thereby reducing the non-stationarity and variance that arise when multiple agents simultaneously adapt their policies. Consequently, the actor can learn scheduling decisions that implicitly account for the impact of its actions on the overall system performance.

%More specifically, during training, the centralized critic updates its parameters by minimizing the value-function loss based on the global state information, whereas each actor updates its policy parameters according to the PPO clipped surrogate objective using the advantages estimated by the critic. This centralized learning mechanism enables the agents to exploit global system information during training without requiring such information during execution. Therefore, the critic effectively serves as a training-time mechanism for coordinating the learning of the distributed queue-level policies rather than as an online decision-making component.

%During the decentralized execution phase, the centralized critic is discarded, and only the trained actor policies are deployed online. Each queue-level agent independently selects its scheduling action based solely on its local observation. Thus, the execution architecture preserves the scalability and distributed nature of TSN queue scheduling, while the centralized training process enables the learned policies to account for the interactions and resource competition among multiple queues.

\subsection{Training and Inference Procedure}
In our model, each TSN queue has a local actor policy $\pi_{\theta_m}$, while a centralized critic $V_\phi$ is used during training. During data collection, each agent observes its local queue state $o_{c,m}^t$ and selects an action according to its policy $\pi_{\theta_i}(a_{c,m}|o_{c,m}^t)$.  The environment then executes the joint actions of all agents, determining the corresponding TSN scheduling decisions, and returns a global team reward $R_c$ and the next system state $s_{t+1}$. These transitions are stored in the buffer $\mathcal{D}$ for subsequent policy updates. After trajectory collection, the centralized critic uses the global system information to estimate the value function, which is used to compute the advantages (i.e., how good each action was relative to expectations) and returns for policy optimization. The actor policies are then updated using the PPO objective, while the centralized critic is updated by minimizing the value-function loss.  During inference, the centralized critic is no longer required, and each queue independently selects its scheduling action using only its local observation and the corresponding trained actor policy.

% \begin{algorithm}[htb!]
% \caption{MAPPO-Based TSN Queue Scheduling}
% \label{alg:mappo_tsn}
% \begin{algorithmic}[1]
% \State Initialize actor $\{\pi_{\theta_i}\}_{i=1}^{N}$, critic $V_{\phi}$, and buffer $\mathcal{D}$
% \State \textit{// Training Phase}
% \For{episode $=1,2,\ldots,L$}
%     \State Initialize TSN system state $s_1$
%     \For{step $t=1,2,\ldots,T$}
%         \State Observe each local queue state $o_{c,i}^t$
%         \State Each agent $i$ executes action as $\pi_{\theta_i}(a_{c,i}|o_{c,i}^t)$
%         \State Observe reward $R_c$ and next state $s_{t+1}$
%     \EndFor
%         \State Store $(s_t,\mathbf{o_{c,i}}_t,\mathbf{a_{c,i}}_t,r_t)$ in $\mathcal{D}$
%     \EndFor
%     \State Compute advantages and returns using the critic
%     \For{epoch $k=1,2,\ldots,K$}
%         \State Update each actor $\pi_{\theta_i}$ using the PPO objective
%         \State Update centralized critic $V_{\phi}$
%     \EndFor
% \State \textit{// Inference Phase}
% \For{each time slot in cycle $c$}
%     \For{agent $i=1,2,\ldots,N$}
%         \State Observe local queue state $o_{c,i}^i$
%         \State Select action $\pi_{\theta_i}(a_{c,i}|o_{c,i}^t)$
%     \EndFor
%     \State Execute joint action
% \EndFor
% \end{algorithmic}
% \label{alg:mappo}
% \end{algorithm}

\section{Performance Evaluation}

\subsection{Experimental Setup}

We use vehicular mobility data collected on the Smart Highway testbed~\cite{bello2026towards}, providing vehicle counts and average speeds for different highway segments. The TSN network has eight queues supporting five XR applications, categorized as AR or SeAR, with traffic modeled using Johnson’s $S_U$ distribution~\cite{morin2023extended}. As shown in Table~\ref{tab:parameters}, queues $q_0$, $q_1$, $q_3$, $q_4$, and $q_7$ handle traditional AR traffic, while the others support different SeAR resolutions with a 2 ms base deadline. All deadlines are scaled by $\alpha \in [0.50,0.90]$, reducing the MEC delay budget from 90\% at low speeds ($v_{avg}^{\min} = 10)$ to 50\% at high speeds ($v_{avg}^{\max} = 120$) to account for speed-dependent latency deadlines. Table~\ref{tab:parameters} also reports the MAPPO hyperparameters. Each episode contains 150 steps, with the number of connected vehicles and their average speed sampled from the dataset at its start. Average speed and vehicle count are clipped to 10--120 and 1--100, respectively. Clients are then distributed among queues using a uniform multinomial distribution, with at least one active user per queue. We set the reward penalty weight to $\lambda=1.0$. Results are reported as the mean and standard error across three random seeds.

\begin{table}[htb]
\centering
\caption{VEC and MAPPO parameters}
\label{tab:parameters}
\fontsize{7.8}{8}\selectfont
\begin{tabular}{@{}cccc@{}}
\toprule
\textbf{VEC}                                  & \textbf{Value}               & \textbf{MAPPO parameter}             & \textbf{Value}            \\ \midrule
Network speed (Gbps)                 & 10                  & Learning rate               & $10^{-4}$        \\
Max packet size (B)                  & 1428                & gamma                       & 0.99             \\
Number of vehicles                   & 1--100              & Episode length              & 150              \\
Average speed (km/h)                 & 10--120              & PPO epochs                  & 5                \\ \midrule
\multicolumn{2}{c}{\textbf{AR setting/Base deadline}}               & 
\multicolumn{2}{c}{\textbf{SeAR setting/Base deadline}} \\
\midrule

\multicolumn{2}{c}{q0, q3: 3840$\times$1920@90 Hz / 11.11 ms} & 
\multicolumn{2}{c}{q2: 1920$\times$720@60 Hz / 2 ms}  \\
\multicolumn{2}{c}{q1, q4, q7: 3840$\times$1920@72 Hz / 13.80 ms} & 
\multicolumn{2}{c}{q5: 1280$\times$480@60 Hz / 2 ms} \\
\multicolumn{2}{c}{} & 
\multicolumn{2}{c}{q6: 2560$\times$960@60 Hz / 2 ms} \\
\bottomrule
\end{tabular}
\end{table}
\textit{Baselines:} To evaluate the performance of our MAPPO approach, we compare it against a set of baselines, including alternative MARL architectures, single-agent algorithms, and a heuristic-based scheduling policy, as described below:

\begin{itemize}
    \item \textbf{MARL:} Multi-agent transformer (MAT)~\cite{wen2022multi} and heterogeneous-agent proximal policy optimization (HAPPO)~\cite{kuba2021trust}, where we also model each queue as an independent agent.
    \item \textbf{Single-agent RL: } PPO and advantage actor-critic (A2C) for centralized control, where a single agent has control over all queues.
    % \item \textbf{Backlog-based heuristic: } A non-learning policy that prioritizes queues with larger backlogs by granting them longer time slots. The time slot for queue $i$ is defined as $t_i = \frac{B_{c,i}}{\sum_{j \in Q} B_{c,j}} \cdot \Delta t_{\text{cycle}}$.
    \item \textbf{Urgency-based heuristic: } A non-learning policy that prioritizes queues approaching their deadlines based on the urgency rate. The time slot allocation is given by $t_i = \frac{d_{c,i}}{\sum_{m=0}^{M-1} d_{c,m}} \cdot \Delta t_{\text{cycle}}$.
\end{itemize}

\subsection{Experimental Results}

\begin{figure}[htb!]
	\centering
    \includegraphics[width=0.8\columnwidth]{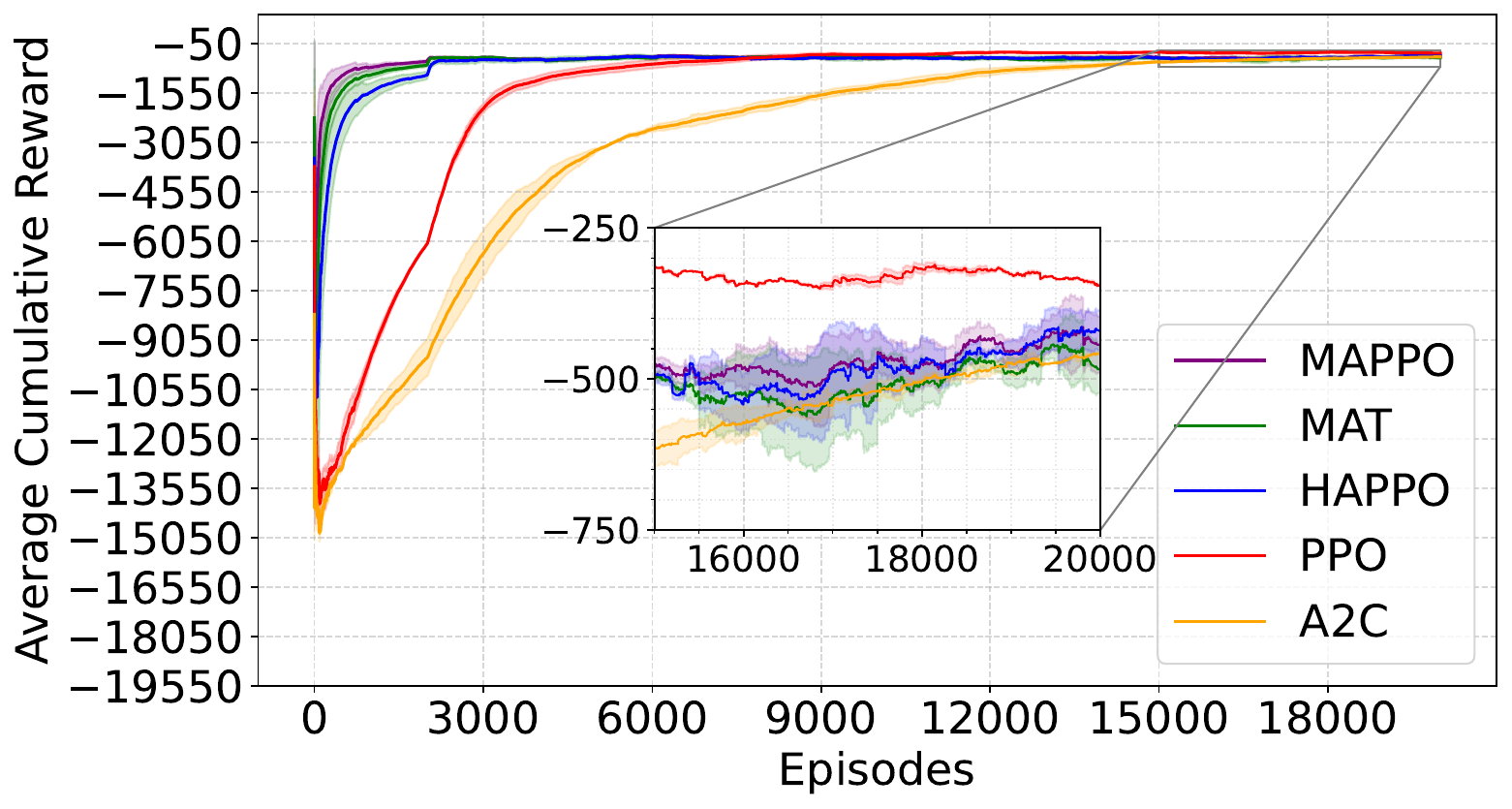}
    \caption{Average cumulative reward.}    
    \label{fig:learning_performance}
\end{figure}

\textit{Learning performance:} As shown in fig.~\ref{fig:learning_performance}, MARL methods converge faster than single-agent methods, reaching stable reward levels in fewer training steps, whereas PPO and A2C require longer training to converge. 
By learning from multiple agents, MARL methods can better capture the impact of queue-level decisions on overall network performance, potentially improving convergence over single-agent methods. Nevertheless, MAPPO converges slightly faster than MAT and HAPPO, likely because it jointly updates all agents using a shared objective, whereas MAT and HAPPO rely on sequential mechanisms that introduce additional dependencies during learning. Despite faster MARL convergence, PPO achieves the highest and most stable final reward. However, higher training rewards do not necessarily imply better network performance under heterogeneous queue deadlines, as shown next.

%\carvalho{One possible explanation is that a single-agent PPO policy may learn a good overall strategy but have difficulty learning the interactions and trade-offs among individual queues.}

\begin{figure}[htb!]
	\centering
    \includegraphics[width=0.9\columnwidth]{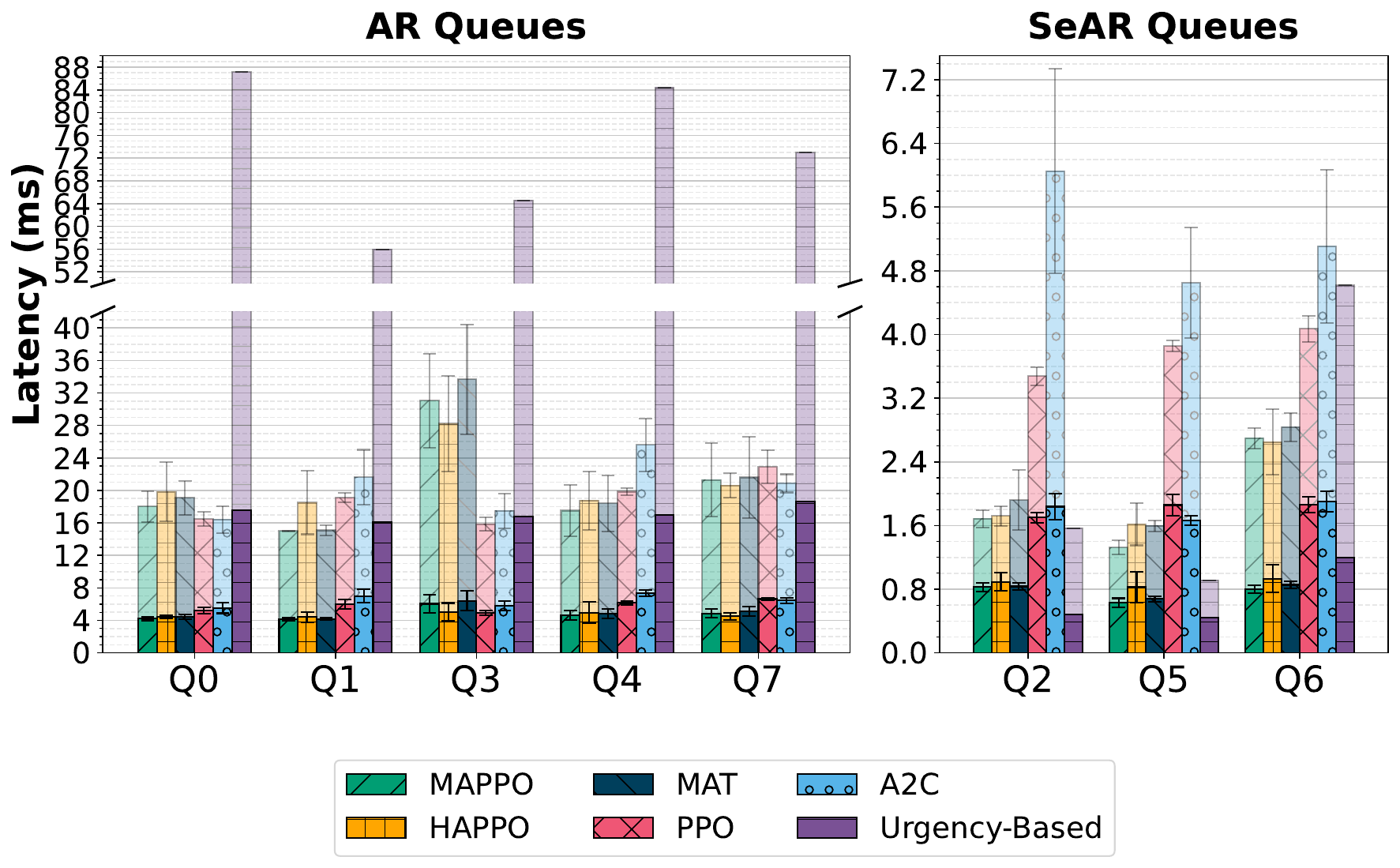}
    \caption{Average (solid fill) and 99th-percentile (transparent fill) PDU-level service latency.}
    \label{fig:pdu_latency}
\end{figure}

\textit{Service Latency Analysis:} Fig.~\ref{fig:pdu_latency} presents the average and $p99$ PDU-level service latency for each queue. As observed, MARL methods achieve lower latency across all queues, reducing the average PDU latency by at least 19.5\% compared with A2C and 14.9\% compared with PPO for the AR queues. Similarly, MARL methods reduce worst-case ($p99$) latency, with reductions of up to 38.6\% compared with the single-agent methods. This performance gap is even more pronounced for SeAR queues, where MARL methods reduce average and $p99$ latency by up to 66.2\% and 68.3\%, respectively, compared with single-agent methods. 

\noindent The results suggest that single-agent policies do not fully capture the interactions and resource trade-offs among queues with heterogeneous requirements, which may explain why PPO achieves higher training rewards than the MARL methods. In particular, PPO and A2C appear to prioritize AR queues, whose higher traffic intensity contributes more frequently to the reward, while postponing service for lower-demand SeAR queues. Although this behavior can increase the number of satisfied AR deadlines and thus the training reward, it may degrade overall network performance by increasing SeAR delay. In contrast, MAPPO, HAPPO, and MAT explicitly model the scheduling problem in a multi-agent setting, enabling more coordinated resource allocation across queues. Consequently, they achieve better execution performance despite lower cumulative training rewards. Nevertheless, MAPPO and MAT slightly outperform HAPPO, exhibiting lower variance across the queues, as indicated by the standard deviation. This difference may be attributed to the distinct coordination mechanisms employed by each method. HAPPO updates the agents sequentially using a randomly shuffled update order, which can introduce variability in the learning process. In addition, urgency-based heuristics generally outperform all learning-based methods for SeAR queues, reducing latency by at least 29.1\%. However, this gain comes at the cost of higher latency in AR queues, with average latency increasing by up to 315\% and worst-case latency reaching 80 ms, corresponding to an 373\% increase compared with learning-based methods.

\begin{figure}[htb!]
	\centering
    \includegraphics[width=0.9\columnwidth]{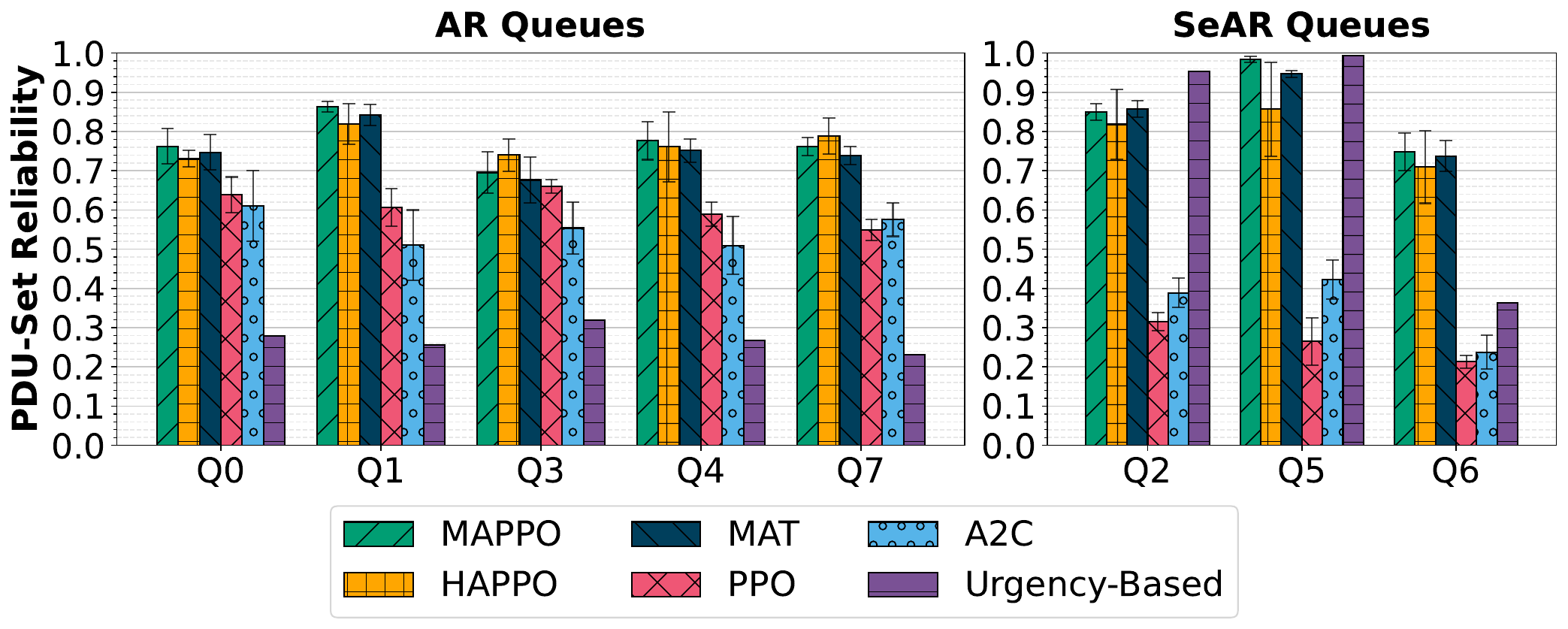}
    \caption{PDU-set reliability (deadline satisfaction).}    
    \label{fig:reliability}
\end{figure}

\textit{PDU-Set Reliability:} Fig.~\ref{fig:reliability} shows the PDU-sets deadline satisfaction rate (i.e., reliability). To achieve high reliability, scheduling policies must ensure that all PDUs belonging to a PDU-set are delivered within its dynamic deadline, as designed by eq.~\ref{eq:reliability}. Accordingly, the reliability results are consistent with the latency analysis, with MARL methods achieving higher reliability across the AR queues, exceeding 70\% compared with approximately 60\% for single-agent methods, corresponding to a 16.66\% improvement. As observed previously, the urgency-based method tends to prioritize SeAR queues due to their stricter deadlines. This urgency-driven approach explains its higher reliability for SeAR traffic, exceeding 90\% for SeAR queues, particularly for $q_{2}$ and $q_{5}$. Despite this improvement, they perform poorly across the AR queues, achieving reliability of only around 30\%. 

Finally, Table~\ref{tab:inference_time} presents the inference time of the learning-based methods on an RTX 4060 Ti. MAPPO and the single-agent methods exhibit lower inference times, whereas HAPPO and MAT exceed the scheduling cycle by more than 5 and 10 times, respectively. This gap results from differences in action generation: MAPPO and single-agent methods generate actions in a single forward pass, while MAT generates them sequentially through an autoregressive policy. HAPPO also incurs higher inference time due to its separate actor network for each agent.
\begin{table}[htb]
\caption{Inference Time (ms)}
\centering
\begin{tabular}{@{}cccc@{}}
\toprule
\textbf{Method} & \textbf{Inference Time} & \textbf{Method} & \textbf{Inference Time} \\ \midrule
MAPPO  & 0.69 $\pm$ 0.18        & MAT    & 10.90 $\pm$ 1.29       \\
HAPPO  & 5.48 $\pm$ 0.61        & PPO    & 0.78 $\pm$ 0.22       \\
A2C    & 0.79 $\pm$ 0.27        &        &                     \\ \bottomrule
\end{tabular}
\label{tab:inference_time}
\end{table}

\noindent Therefore, although HAPPO and MAT achieve competitive performance, their higher inference latency may limit real-time deployment under a 1 ms control cycle. In contrast, MAPPO achieves comparable performance with significantly lower inference latency.

% \begin{table}[htb!]
%     \centering
%     \caption{Inference time}
%     \label{tab:inference_time}
%     \begin{tabular}{lc}
%         \toprule
%         \textbf{Model} & \textbf{Inference Time (ms)} \\
%         \midrule
%         MAPPO & $0.72 \pm 2.11$ \\
%         HAPPO & $5.50 \pm 2.00$ \\
%         MAT   & $10.93 \pm 2.81$ \\
%         PPO   & $0.80 \pm 1.48$ \\
%         A2C   & $0.81 \pm 1.41$ \\
%         \bottomrule
%     \end{tabular}
% \end{table}

\section{Conclusion}
This paper addresses co-located XR traffic scheduling in VEC-TSN integration by considering vehicle speed-dependent latency deadlines. We leverage MAPPO to model queue-level interdependencies under dynamic deadlines, using a centralized critic to learn cooperative scheduling policies. Simulation results show that MAPPO outperforms the considered baselines by reducing latency, improving PDU-set reliability, and maintaining low inference time, making it suitable for real-time scheduling. In future work, we intend to evaluate the impact of wireless networks on the proposed framework.
\label{sec:conclusion}

\section*{Acknowledgment}
This work was financed by the Coordenação de Aperfeiçoamento de Pessoal de Nível Superior - Brasil (CAPES) - Finance Code 001, CNPq (funding agency from the Brazilian federal government), FAPEMIG (Minas Gerais State Funding Agency), and São Paulo Research Foundation (FAPESP) with Brazilian Internet Steering Committee (CGI.br), grants 2018/23097-3.

\bibliographystyle{IEEEtran}
\bibliography{references}

\end{document}